\documentclass{PoS}

\usepackage{amsmath}
\usepackage{placeins}
\usepackage{array}

\title{The resummation of the low-phistar domain of Z production}

\ShortTitle{The resummation of the low-phistar domain of Z production}

\author{\speaker{Lee TOMLINSON}%
         \thanks{I give thanks to my collaborators on this work: A. Banfi, M. Dasgupta and S. Marzani. This work was supported by the UK Science and Technology Facilities Council (STFC).}\\
        ( University of Manchester ( GB))\\
        E-mail: \email{lee.tomlinson@cern.ch}}

\abstract{The presence of large logarithms in QCD corrections to observables spoils the validity of a calculation truncated at finite order and calls for an all-orders approach. The transverse momentum, $Q_T$, spectrum of massive lepton pairs, produced in hadron colliders by the Drell-Yan mechanism, has received a great deal of attention in electroweak phenomenology. We present and discuss a next-to-next-to-leading log (NNLL) resummed calculation of a related observable, namely $\phi^*$ (phistar), that was recently introduced because of its distinct experimental advantages, but which is nonetheless sensitive to similar physics: soft-collinear gluon emission in the initial state. We also present various comparisons to collision data at the Tevatron and the LHC.}

\FullConference{XXI International Workshop on Deep-Inelastic Scattering and Related Subjects\\
   22-26 April, 2013\\
    Marseilles, France}

\begin{document}

\section{Introduction}

The production of massive lepton pairs at hadron-hadron colliders via the \emph{Drell-Yan} mechanism \cite{Drell:1970wh} is one of the most studied processes in particle physics phenomenology on account of its several remarkable properties, most notably its clean experimental signature and, from a theoretical perspective, its proven factorisation in the context of the QCD factorisation theorem. This makes it particularly suited to detailed phenomenology of QCD corrections. 
In particular, the transverse momentum spectrum, $Q_T$, of the associate gauge boson is of particular importance experimentally in terms of the mass measurement of the W boson, and theoretically as a probe of the effects of soft-collinear QCD radiation from the incoming partons, to give two concrete examples. These proceedings review a theoretical calculation, designed to capture the soft-collinear effects, and the associated phenomenology \cite{Banfi:2009dy,Banfi:2011dx,Banfi:2012du,Banfi:2011dm}.

In order for a non-zero $Q_T$ spectrum of the EW gauge boson to arise, it must either be produced from partons whose net transverse momenta were initially non-zero due to their \emph{intrinsic} motion within the proton, or which obtained non-zero transverse momenta by recoiling against some emission. The scales at which such non-zero $Q_T$ could occur define three distinct r\'{e}gimes one must consider: $Q_T\sim M\sim M_\textrm{Z}$, $\Lambda_\textrm{QCD}\ll Q_T\ll M$ and $\Lambda_\textrm{QCD}\sim Q_T$. Similar r\'{e}gimes exist for the $\phi^*$ (phistar) observable to be defined below. The importance of these distinctions lies with the calculability of the observable. As $\Lambda_\textrm{QCD}\sim Q_T$, the process falls out of the jurisdiction of a perturbative calculation. However, for $Q_T\sim M$ a perturbative calculation truncated at finite order describes the physics well.

\begin{figure}
\centering
\begin{tabular}{m{0.45\columnwidth}m{0.5\columnwidth}}
\includegraphics[width=0.45\columnwidth]{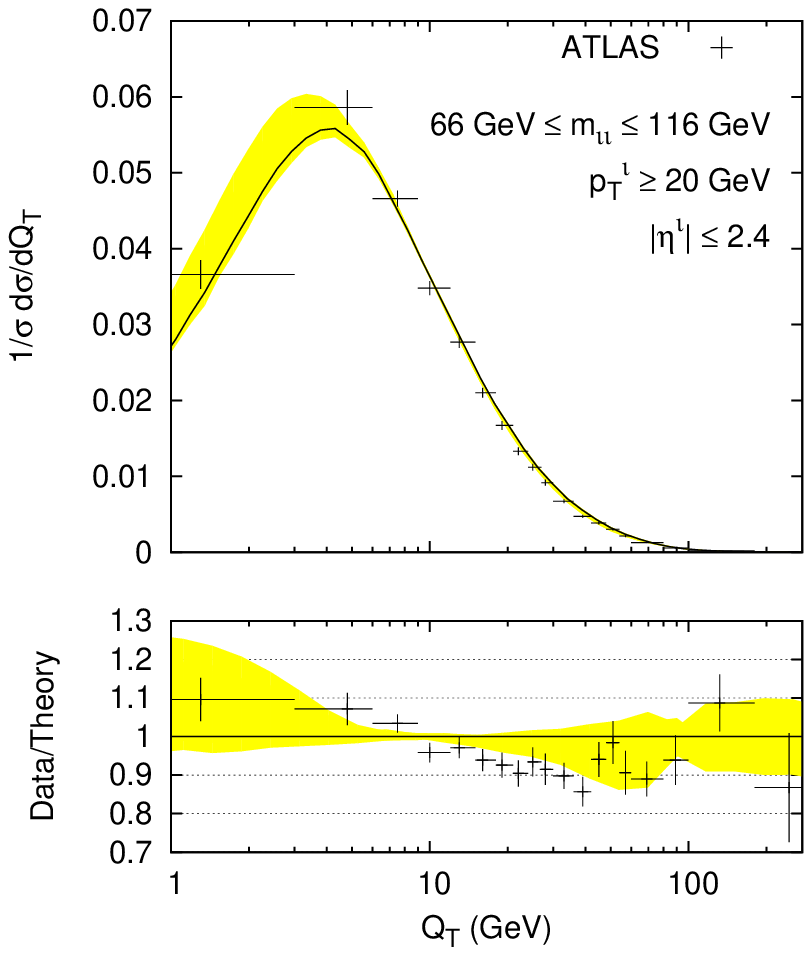}&
\includegraphics[width=0.5\columnwidth]{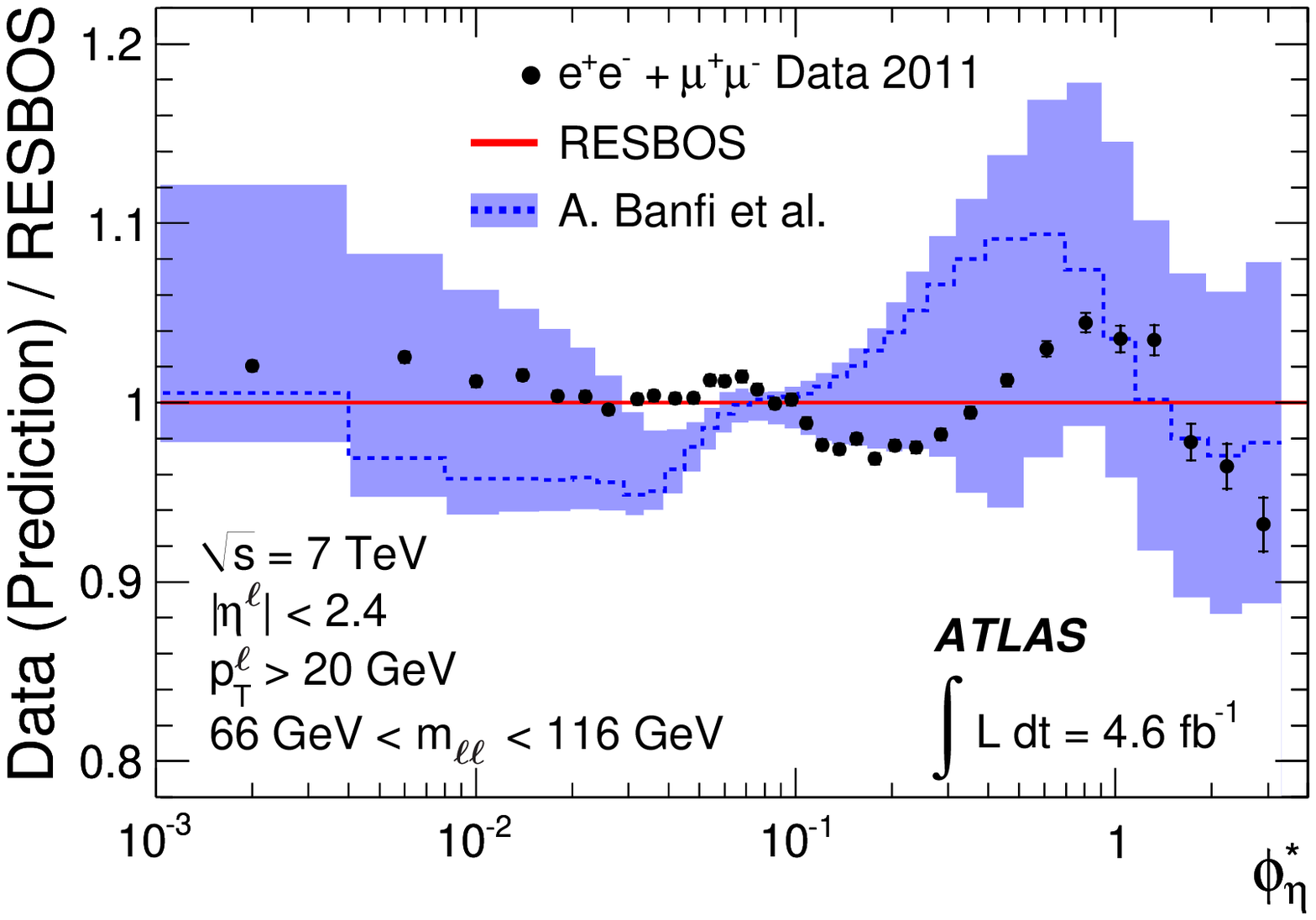}
\label{fig:phistar}
\end{tabular}
\caption{Left: A comparison of the computed (NNLL+NLO) distribution to ATLAS data \cite{Aad:2012wfa} for the $Q_T$ spectrum of the Z boson in Drell-Yan events, as a means of validating the resummation. Right: A prediction of $\phi^*$ for ATLAS at $\sqrt{s}=7$ TeV, shown by the blue dashed line, compared to recent ATLAS data. The yellow (left) and blue (right) theoretical uncertainty bands are obtained by making perturbative scale variations independently.}
\label{fig:atlas}
\end{figure}

This calculation becomes divergent at low values of $Q_T$, which is (double-) logarithmic in the ratio of the two disparate physical scales $Q_T$ and $M$. The presence of these logarithmic enhancements persists to all orders, rendering the \emph{effective} perturbative expansion parameter potentially greater than unity---i.e. $\alpha_s \ln^2(M/Q_T) \gtrsim 1$---spoiling the formal validity of a perturbative calculation truncated at finite order. Such a calculation is treated using resummation, whereby the calculation is restructured and entire classes of logarithms are summed to all orders. The phenomenology reported herein makes use of a state-of-the-art resummed calculation, accurate to next-to-next-to-leading log (NNLL), which has most recently been compared with ATLAS data \cite{Aad:2012wfa} at $\sqrt{s}=7$ TeV, for both the $Q_T$ and $\phi^*$ observables, as shown in Fig. \ref{fig:atlas}.

\begin{figure}
\centering
\includegraphics[width=0.45\columnwidth]{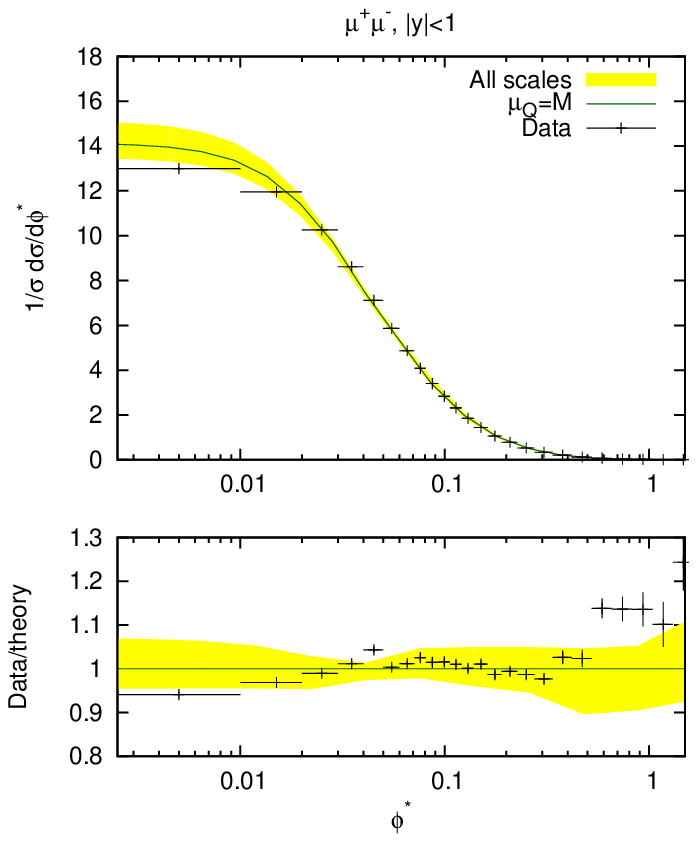}
\includegraphics[width=0.45\columnwidth]{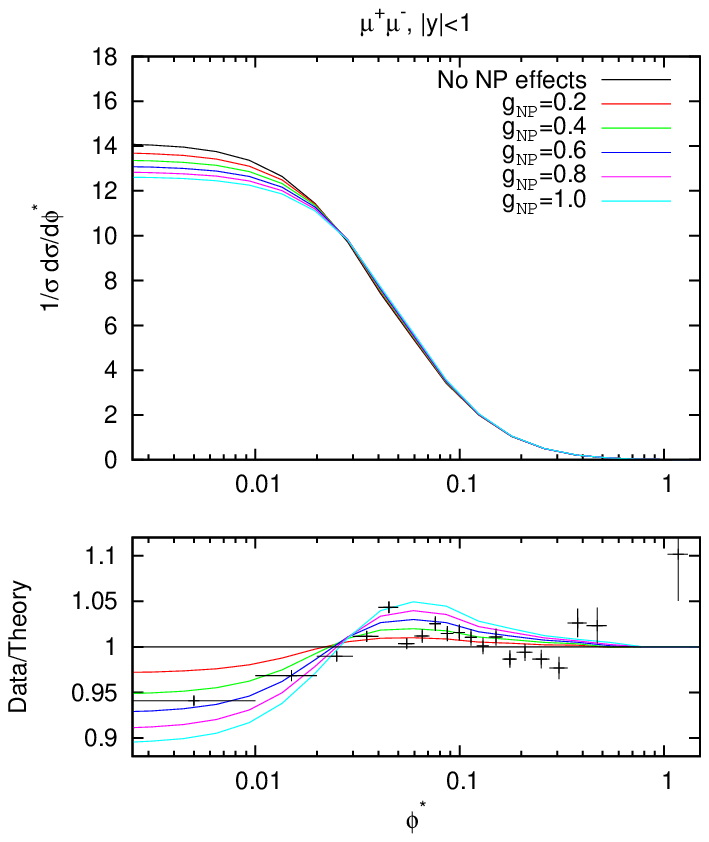}
\label{fig:d0results}
\caption{Left: A comparison of the computed (NNLL+NLO) distribution to D\O\ data \cite{Abazov:2010mk} in a central rapidity region, $|y|<1$, for $\phi^*$. The yellow theoretical uncertainty band is obtained by making perturbative scale variations independently. Right: The same comparison but with a non-perturbative (NP) model parameterised by $g_\textrm{NP}$ included. The choice of $g_\textrm{NP}\approx0.6$ gives the best agreement with data.}
\label{fig:d0}
\end{figure}

Equipped with a sound theoretical calculation of the $\phi^*$ spectrum of the Z boson at hadron colliders, and given the impressive experimental uncertainty to which this observable has been measured, one may begin to look for differences between data and theory. The calculation described herein is perturbative in nature---i.e. no non-perturbative (NP) models have been used, for instance, to model intrinsic parton $k_T$, etc.---and, with that in mind, one may be able to address the question of the need for specific NP models. Indeed, the study of a simple NP model, in the form of a Gaussian smearing term, has been carried out on top of the perturbative calculation we present in order to test its effects on the $\phi^*$ spectrum. This is shown in Fig. \ref{fig:d0}, alongside the purely perturbative prediction with scale uncertainties included, both in comparison with D\O\ data \cite{Abazov:2010mk}. Within the current perturbative scale uncertainties, however, the actual \emph{need} of a NP model remains inconclusive for this study.

\newpage
\section{The $\phi^*$ observable}

The $\phi^*$ observable, quite recently introduced \cite{Banfi:2010cf}, proves to be experimentally better-determined than $Q_T$ by virtue of its entirely angular construction, rendering it largely immune to momentum resolution inefficiencies that affect the $Q_T$ measurement. The $\phi^*$ observable may be defined according to
\begin{equation}
\phi^*:=\tan\left(\frac{\phi_\textrm{acop}}{2}\right)\,\sin\theta^*,
\end{equation}
where the asterisk denotes the frame in which the leptons are produced back-to-back \emph{longitudinally}. In this frame, $\theta^*$ is the angle made by the lepton(s) with respect to the beam axis. This observable is clearly sensitive to similar physics as the familiar $Q_T$ because it measures the `deviation from back-to-backness' (acoplanarity) in the transverse plane, which is produced by the same mechanism that generates non-zero $Q_T$, namely recoil against other emission.

The importance of the distinct scale r\'{e}gimes has already been discussed. One would like to compute a distribution which is valid across a wide range of $\phi^*$ and, as such, a smooth matching prescription is employed, where
\begin{equation}
\left(\frac{d\sigma}{d\phi^*}\right)_\textrm{matched}=
\left(\frac{d\sigma}{d\phi^*}\right)_\textrm{resummed}+
\left(\frac{d\sigma}{d\phi^*}\right)_\textrm{fixed-order}-
\left(\frac{d\sigma}{d\phi^*}\right)_\textrm{expanded}.
\end{equation}
The term denoted `expanded' is the resummed calculation expanded to the same perturbative order to which the fixed-order calculation is performed, in order to subtract the double-counting. This work makes use of a next-to-leading order (NLO\footnote{NLO here refers to the \emph{distribution}, not the inclusive cross-section.}) calculation provided by MCFM \cite{Campbell:2002tg} for the distributions, that is $\mathcal{O}(\alpha^2_\textrm{s})$.

The resummed distribution has the following form:
\begin{equation}
\begin{split}
\frac{d\sigma}{d\phi^*}(\phi^*,M,\cos\theta^*,y)=\frac{\pi\,\alpha^2}{s\,N_c}
\int^\infty_0&db\,M\,\cos\theta^*(b\,M\,\phi^*)\,e^{-R(\bar{b},M,\mu_Q,\mu_R)}\\
&\times\Sigma(x_1,x_2,\cos\theta^*,b,M,\mu_Q,\mu_R,\mu_F),
\end{split}
\end{equation}
where
\begin{equation}
x_{1,2}=\frac{M}{\sqrt{s}}e^{\pm y}\quad\textrm{and}\quad
\bar{b}=\frac{b\,e^{\gamma_\textrm{E}}}{2}.
\end{equation}
In this expression, $\Sigma$ is essentially the Born calculation, but with appropriate coefficient functions, and with the PDFs evaluated at the scale $\mu_F/(\bar{b}\mu_Q)$, automatically resumming the collinear logs according to DGLAP. The so-called \emph{radiator}, $R(\bar{b})$, encapsulates all the logarithms arising from soft-collinear emission that we wish to resum, and is given by the perturbative expression
\begin{equation}
R(\bar{b},M,\mu_Q,\mu_R)=L\,g^{(1)}(\alpha_s\,L)+g^{(2)}\left(\alpha_s\,L,\frac{M}{\mu_Q},\frac{\mu_Q}{\mu_R}\right)+\frac{\alpha_s}{\pi}\,g^{(3)}\left(\alpha_s\,L,\frac{M}{\mu_Q},\frac{\mu_Q}{\mu_R}\right),
\end{equation}
where $L=\log\big(\bar{b}^2\,\mu_Q^2\big)$ and $\alpha_s=\alpha_s(\mu_R)$ in the $\overline{\textrm{MS}}$ scheme. The scales $\mu_Q$, $\mu_R$ and $\mu_F$ are the resummation, renormalisation and factorisation scales, respectively. The explicit expressions for $g^{(i)}$ are summarised in the appendices of \cite{Banfi:2011dm}.

\FloatBarrier

\section{Conclusions}

The theoretical predictions for $\phi^*$ presented here agree with data from D\O\ and ATLAS within theoretical uncertainties, which are derived from a reasonable variation of the three perturbative scales in the problem about the nominal value $M$. The dominant uncertainty is due to the resummation scale, $\mu_Q$, which swamps all other uncertainties in the low-$\phi^*$ region we are most interested in. Within these scale uncertainties, no conclusive statement can be made about the need for a NP model, since any discrepancy with data can, at present, be mitigated by an appropriate choice of scales. The exploration of more extreme kinematic r\'{e}gimes may begin to challenge standard $Q_T$ resummation within current uncertainties, however, providing the opportunity to study $x$-dependent non-perturbative models in the radiator, for example. The desire to have a reduced theoretical uncertainty will nonetheless form part of future work.

\end{document}